\documentclass{aa}
\usepackage{txfonts}
\usepackage{graphicx}
\usepackage{natbib}
\bibpunct{(}{)}{;}{a}{}{,}

\begin{document}

\title{On the formation of terrestrial planets in hot--Jupiter systems}
\author{Martyn J. Fogg \& Richard P. Nelson.}
\institute{Astronomy Unit, Queen Mary, University of London, Mile
End Road, London E1 4NS.\\
\email{M.J.Fogg@qmul.ac.uk, R.P.Nelson@qmul.ac.uk}}
\date{Received/Accepted}

\abstract
{There are numerous extrasolar giant planets which orbit
close to their central stars. These `hot-Jupiters' probably
formed in the outer, cooler regions of their
protoplanetary disks, and migrated inward to $\sim 0.1$ AU.
Since these giant planets must
have migrated through their inner systems at an early time, it is
uncertain whether they could have formed or retained terrestrial
planets.}
{We present a series of calculations aimed at examining
how an inner system of planetesimals/protoplanets,
undergoing terrestrial planet formation, evolves under the
influence of a giant planet undergoing inward type II migration
through the region bounded between 5 -- 0.1 AU.}
{We have previously simulated the effect of gas giant planet
migration on an inner system protoplanet/planetesimal disk using a
N-body code which included gas drag and a prescribed
migration rate. We update our calculations here with an
improved model that incorporates a viscously evolving gas disk,
annular gap and inner--cavity formation due to the gravitational
field of the giant planet, and self--consistent evolution of the giant's
orbit.}
{We find that $\gtrsim$ 60\% of the solids disk survives by being
scattered by the giant planet into external orbits. Planetesimals
are scattered outward almost as efficiently as protoplanets,
resulting in the regeneration of a solids disk where dynamical
friction is strong and terrestrial planet formation is able to
resume. A simulation that was extended for a few Myr after the
migration of the giant planet halted at 0.1 AU, resulted in an
apparently stable planet of $\sim 2$ m$_{\oplus}$ forming in the
habitable zone. Migration--induced mixing of volatile--rich material
from beyond the `snowline' into the inner disk regions means that
terrestrial planets that form there are likely to be water--rich.}
{We predict that hot--Jupiter systems are likely to harbor
water--abundant terrestrial planets in their habitable zones. These
planets may be detected by future planet search missions.}

\keywords{planets and satellites: formation -- methods: N-body
simulations -- astrobiology}
\titlerunning{Growth of terrestrial planets in the
presence of gas giant migration.}
\authorrunning{M.J. Fogg \& R.P. Nelson}

\maketitle

\section{Introduction.}\label{intro}

About one quarter of the extrasolar planetary systems discovered to
date contain a so-called `hot-Jupiter' -- a gas giant planet
orbiting within 0.1 AU of the central star \citep[e.g.][]{butler2}.
It is improbable that these planets formed within such a hot region
of the protoplanetary disk, and is most likely that they originated
further out beyond the nebula snowline and moved inward via type II
migration \citep[e.g][]{lin1,ward2,bryden,nelson1}. Since both giant
planet formation and type II migration require the nebular gas to
still be present, these processes are constrained to occur within
the first few million years of the disk lifetime \citep{haisch},
well within the $\sim$ 30 -- 100~Myr timescale thought to be
required to complete the growth of terrestrial planets
\citep[e.g.][]{chambers2,kleine,halliday}. Thus, these migrating
giant planets must have traversed the inner system, including its
habitable zone, before any planet formation there was complete,
raising the question of what effect such a disturbance would have on
the growth of terrestrial--like planets.

Initially, it was thought reasonable that the inward migration of a
giant planet would be so disruptive of the material it passed
through as to clear the swept zone completely, precluding the
formation of any inner system planets. However, this view was a
conservative assumption, often made in the support of speculative
astrobiological arguments
\citep[e.g.][]{ward1,lineweaver1,lineweaver2}. Agreement as to the
outcome failed to materialize from the first modeling studies of the
process, the conclusions of which varied from the occurrence of
terrestrial planets in hot-Jupiter systems being highly unlikely
\citep{armitage2}, through possible but rare \citep{mandell}, to
commonplace \citep{raymond2}.
None of these studies, however, actually simulated terrestrial planet
formation simultaneously with giant planet migration. Their
disagreement about the likely outcome can be traced to assumptions
made about the timing of giant planet formation and migration.

The first study to model inner system planetary accretion in the
presence of a migrating giant was that of \citet{fogg} (hereafter
referred to as Paper I). This work used N--body simulations to
examine oligarchic and giant--impact growth \citep{kokubo} in a
protoplanet/planetesimal disk based on the minimum mass solar nebula
model of \citet{hayashi}, extending between 0.4 -- 4 AU. Five
scenarios were considered, corresponding to five different ages for
the inner planet forming disk at the point  when a giant planet was
assumed to form at 5 AU and migrate in to 0.1 AU.

In all five of their scenarios, \citet{fogg} found that the majority
of the disk solids survived the passage of the giant planet, either
by being shepherded inward of the giant, or by being scattered by
the giant into excited exterior orbits. This partition of solid
material was shown to vary with the level of dissipative forces
present, which decline with disk maturity, favoring shepherding at
early times and scattering at late times. Within the portion of the
disk compacted inside the increasingly restricted volume interior to
the giant, accretion was found to speed up greatly at late times
resulting typically in a $\sim 3-15~\mathrm{M}_\oplus$ planet
forming inside 0.1~AU. The similarity of these objects to the
recently identified class of `hot-Neptune' planets
\citep{mcarthur,butler1,santos04b,vogt,rivera,bonfils,udry} was
noted and discussed. The fate of the material scattered into
external orbits was not subjected to further calculation, but it was
noted that, although the remaining solids surface density was
reduced from pre-migration-episode values, ample material remained
to provide for the eventual accretion of a set of external
terrestrial planets, including within each system's habitable zone.
\citet{fogg} therefore concluded that the assumption that
hot-Jupiter systems are devoid of inner system terrestrial planets
is probably incorrect, and that planet formation and retention both
interior and exterior to the hot-Jupiter is possible. We note that
similar results, relating to the formation of planets interior to a
migrating giant, have been reported by \citet{zhou}.

In this paper, we extend the model introduced in Paper I by
improving the realism with which gas dynamics is simulated. A 1-D
evolving viscous gas disk model is linked to the N-body code that:
1) allows the gas to deplete over time via viscous accretion onto
the central star; 2) allows an annular gap to form in the vicinity
of the giant planet; 3) includes the creation of an inner cavity due
to dissipation of propagating spiral waves excited by the giant
planet; 4) self-consistently drives the giant inward. Compared to
the unevolving gas disk assumed in Paper I, this new model reduces
the strength of dissipation present in all scenarios, especially in
regions close to the central star and the giant. We examine and
re-interpret the fate of the disk solids under these changed
circumstances and look at whether our hypothesis of hot-Neptune
formation remains robust. We also examine the post--migration
evolution of the outer scattered disk of solids, and find that
terrestrial planets do form in the habitable zone. Another issue we
examine is the extent to which volatile-rich matter, originating
from beyond the snowline, is driven into the inner system and is
mixed into the surviving planet-forming material. We find that
substantial mixing occurs, such that any terrestrial planets that
form are likely to be water--rich bodies hosting deep, global
oceans.

The plan of the paper is as follows.
In Section 2 we outline the additions to our model and the initial
conditions of the simulations; in Section 3 the results are
presented and discussed; in
Section 4 we consider some caveats and future model improvements,
and in Section 5 we offer our conclusions.

\section{Description of the model.}\label{description}

We model planetary accretion using the \emph{Mercury 6}
hybrid-symplectic integrator \citep{chambers1}, run as an $N + N'$
simulation, where we have $N$ protoplanets embedded in a disk of
$N'$ ``super-planetesimals" -- particles that represent an idealized
ensemble of a much larger number of real planetesimals. The
protoplanets (and the giant when it is introduced) interact
gravitationally and can accrete with all the other bodies in the
simulation, whereas the super-planetesimal population is
non-self-interacting. These latter objects however are subject to a
drag force from their motion relative to the nebular gas. A detailed
outline of these aspects of our model is given in Sect.~2.1 \& 2.2
of Paper I and we continue here to discuss the additional features
we have introduced.

\subsection{Improved nebula model.}\label{gasmodel}

In Paper I we assumed a steady state gas disk model with a constant
surface density profile $\propto r^{-1.5}$. The migration rate of
the giant was prescribed from a calculation of the local viscous
evolution timescale. More realistically, the quantity of nebular gas
should decline due to viscous evolution and accretion onto the
central star, progressively depleting the inner disk. The gas
density should also decrease in the vicinity of the giant due to the
generation and dissipation of density waves at Lindblad resonance
positions, clearing an annular gap in a zone where the planetary
tidal torques dominate the intrinsic viscous torques of the disk. A
consistent calculation of the type II migration rate should involve
the back-reaction to these tidal torques.

To account for these processes we model the gas disk by solving
numerically the disk viscous diffusion equation \citep{pringle},
with modifications included to account for the tidal influence of an
embedded giant planet. Such a method has been used previously in
studies that attempt to explain the statistical distribution of
exoplanetary orbits through type II migration and disk dispersal
\citep{trilling,armitage1,alibert}. The simplest technique for
including the effect of the planet is the impulse approximation of
\citet{lin1}, where wave dissipation is assumed to occur close to
the planet. A more sophisticated treatment of the problem is the WKB
approximation \citep{takeuchi} which involves summing the torque
contributions from a series of Lindblad resonances in the disk. The
former technique was adopted in the studies cited above as it
requires considerably less computation and generates comparable
results. We follow this approach here, but in order to include the
effect of non-local dissipation of waves that travel far into the
disk we also include the WKB approximation torques due to the waves
launched at the two innermost and outermost Lindblad resonances.

We assume a MMSN-type protoplanetary disk around a $M_\ast = 1~
\mathrm{M}_\odot$ star \citep{hayashi}. The initial surface density
of solids is:

\begin{equation}\label{solids}
\Sigma_{\mathrm{s}}(r) = f_{\mathrm{neb}} f_{\mathrm{ice}} \Sigma_1
\left( \frac{r}{1\mathrm{AU}}\right)^{-1.5}\ ,
\end{equation}
where $r$ is radial distance from the central star,
$f_{\mathrm{neb}} $ is a nebular mass scaling factor, $\Sigma_1 = 7\
\mbox{g cm}^{-2} $ and the ice condensation coefficient
$f_{\mathrm{ice}}=1$ for $a< 2.7\ \mbox{AU} $ (the distance chosen
for the nebula `snowline') and $f_{\mathrm{ice}}=4.2$ for $a \ge
2.7\ \mbox{AU} $. As in Paper I, we set $f_{\mathrm{neb}} =3$.

\noindent The initial surface density of gas is:

\begin{equation}\label{gas}
\Sigma_{\mathrm{g}}(r) = f_{\mathrm{neb}} f_{\mathrm{gas}} \Sigma_1
\left( \frac{r}{1\mathrm{AU}}\right)^{-1.5}\ ,
\end{equation}
where $f_{\mathrm{gas}}$ is the gas to dust ratio which we take to
be $f_{\mathrm{gas}} = 240$.

\noindent Given a nebular radial temperature profile of $T_\mathrm{neb} =
280(r/\mathrm{AU})^{-1/2}$, the sound speed of a solar composition
gas in cgs units is:

\begin{equation}\label{sound}
c_\mathrm{s} = 9.9\times 10^4 \left(
\frac{T_\mathrm{neb}}{280~\mathrm{K}} \right)^\frac{1}{2}\ ,
\end{equation}
and the gas scale height is:

\begin{equation}\label{scaleheight}
h = 0.047 \left( \frac{r}{1~\mathrm{AU}} \right)^\frac{5}{4}
\mathrm{AU}\ .
\end{equation}
Since the kinematic viscosity in an alpha-disk model is $\nu =
\alpha h^2 \Omega$, where $\Omega$ is the local Keplerian angular
velocity, we take the turbulent shear viscosity function of the disk
to be:

\begin{equation}\label{viscosity}
\nu = 9.84 \times 10^{16} \alpha \left( \frac{r}{1~\mathrm{AU}}
\right)\ ,
\end{equation}
and the viscous evolution time:

\begin{equation}\label{viscoustime}
\tau_\nu = \frac{2}{3} \left( \frac{r}{h} \right)^2 ( \alpha
\Omega)^{-1} \approx 10^{-4} \alpha^{-1} r \approx 48~\alpha^{-1}
\left( \frac{r}{1~\mathrm{AU}} \right)~~\mathrm{yr}\ .
\end{equation}
In all the models presented here we assume a disk alpha viscosity of
$\alpha = 2\times10^{-3}$, giving $\tau_\nu(5~\mathrm{AU}) \approx
120\,000$ years.

We solve the diffusion equation for $\Sigma_\mathrm{g}(r)$ in the form:

\begin{equation}\label{sigmaevol}
\frac{\partial \Sigma_\mathrm{g}}{\partial t} = \frac{1}{r}
\frac{\partial}{\partial r} \left[ 3r^\frac{1}{2}
\frac{\partial}{\partial r} \left(\nu \Sigma_\mathrm{g}
r^\frac{1}{2}\right) - \frac{2 \Lambda \Sigma_\mathrm{g}
r^\frac{3}{2}}{(GM_\ast)^\frac{1}{2}} - \frac{T r^\frac{1}{2}}{3\pi
(GM_\ast)^\frac{1}{2}} \right]\ ,
\end{equation}
where the first term in square brackets describes the diffusion of
gas under the action of internal viscous torques \citep{pringle};
the second term describes the impulse approximation of the local
tidal interaction of the planet with the disk, with $\Lambda$ being
the specific torque exerted by the planet \citep{lin1}; and the
third term, which derives from the WKB approximation
\citep{takeuchi}, is included to account for more distant angular
momentum transfer via the damping of waves launched from the
innermost and outermost two Lindblad resonances, with $T$ being a
summation of the torque densities exerted by these waves. Because
their launch sites stand off a substantial distance from the planet,
these waves are expected to be linear and not damped locally in the
disk. Their angular momentum content is therefore deposited in the
disk through viscous damping as they propagate.

The exchange of angular momentum between the planet and disk leads
to a radial migration of the planet at a rate:

\begin{equation}\label{adot}
\frac{\mathrm{d}a}{\mathrm{d}t} = - \left( \frac{a}{GM_\ast}
\right)^\frac{1}{2} \frac{1}{m_\mathrm{p}} \left[ 4\pi
\int^{r_\mathrm{out}}_{r_\mathrm{in}} r \Lambda \Sigma_\mathrm{g}
\mathrm{d}r + 2\int^{r_\mathrm{out}}_{r_\mathrm{in}} T \mathrm{d}r
\right]\ ,
\end{equation}
where $m_\mathrm{p}$ is the mass of the planet, $a$ is its
semi-major axis and $r_\mathrm{in}$ and $r_\mathrm{out}$ are the
inner and outer boundaries of the disk respectively.

The rate of specific angular momentum transfer to the disk in the
impulse approximation is given by \citet{lin1} as:

\begin{equation}\label{lambda}
\Lambda = \mathrm{sign} \left(r - a \right)~q^2~\frac{G M_\ast}{2r}
\left( \frac{r}{|\Delta_\mathrm{p}|} \right)^4\ ,
\end{equation}
where $q = m_\mathrm{p} / M_\ast$, and $\Delta_\mathrm{p} =
\mathrm{max} (h, |r - a|)$.

The total torque density exerted on the disk in the WKB
approximation via the damping of waves excited by the planet is
\citep{takeuchi}:

\begin{equation}\label{tottorque}
T(r) = \sum_m T_m(r)\ ,
\end{equation}
where $m$ is the mode number of the $m\mathrm{th}$ order Lindblad
resonance at:

\begin{equation}\label{Lindblad}
r_\mathrm{L} = \left( 1 \mp \frac{1}{m} \right)^\frac{2}{3} a \ ,
\end{equation}
where use of the minus sign gives the radial distances of the inner
resonances $r_\mathrm{IL}$ and the plus sign those of the outer
resonances $r_\mathrm{OL}$. Since only the innermost and outermost
two Lindblad resonances are accounted for here, we take $m = 2,3$
for the resonance positions interior and $m = 1,2$ for those
exterior to the planet.

The torque density is calculated from the radial gradient of the
angular momentum flux $F_m(r)$:

\begin{equation}\label{torque}
T_m(r) = - \frac{\mathrm{d}F_m(r)}{\mathrm{d}r}\ ,
\end{equation}
which is given in \citet{takeuchi} as:

\begin{eqnarray}\label{amflux}
F_m(r) =  F_{m0}\exp \left[ - \int^r_{r_\mathrm{L}} \left\{ \zeta +
\left( \frac{4}{3} + \frac{\kappa^2}{m^2(\Omega -
\Omega_\mathrm{p})^2} \right) \nu \right\} \right. \nonumber
\\
\left. \frac{m(\Omega_\mathrm{p} - \Omega)}{c_\mathrm{s}^2} k
\mathrm{d}\widetilde{r} \right]\ ,
\end{eqnarray}
where $F_m(r) = 0$ for $r_\mathrm{IL} < r < r_\mathrm{OL}$, $\zeta$
is the bulk viscosity (set here to zero), $\kappa = \Omega$ for a
Keplerian disk, $\Omega_\mathrm{p}$ is the angular velocity of the
planet and $k(r)$ is the radial wavenumber:

\begin{equation}\label{wavenumber}
k(r) = \left[ \frac{m^2(\Omega - \Omega_\mathrm{p})^2 -
\kappa^2}{c_\mathrm{s}^2} \right]^\frac{1}{2}\ .
\end{equation}
\citet{takeuchi} give this approximation for the angular momentum
flux originating at a given $m\mathrm{th}$ order resonance:

\begin{eqnarray}\label{fluxatm}
\lefteqn{F_{m0} =  \frac{4}{3} m^2 f_\mathrm{c}
\Sigma_\mathrm{g}(r_\mathrm{L}) \left( \frac{G m_\mathrm{p}}{a
\Omega_\mathrm{p}} \right)^2} \nonumber \\
& &\left[2\mathrm{K_0}\left(\frac{2}{3} \right) +
\mathrm{K_1}\left(\frac{2}{3} \right) - \frac{\pi}{2}
\delta_{m,1}(2\pm1) \right]^2\ ,
\end{eqnarray}
where $\mathrm{K}_0$ and $\mathrm{K}_1$ are modified Bessel
functions, $\delta_{m,1}$ is the Kronecker delta function, and the
upper component of the $\pm$ is used for the inner and the lower
component for the outer resonances respectively.

The parameter $f_\mathrm{c}$ is a torque cutoff function
\citep{artymowicz} given by:

\begin{equation}\label{cutoff}
f_\mathrm{c} = \frac{1}{H(1 + 4 \xi^2)} \left[ \frac{2H
\mathrm{K_0}(2H/3) + \mathrm{K_1}(2H/3)}{2\mathrm{K_0}(2/3) +
\mathrm{K_1}(2/3)} \right]^2\ ,
\end{equation}
where $H = (1 + \xi^2)^{1/2}$ and $\xi =
m(c_\mathrm{s}/a\Omega_\mathrm{p})$.

The evolution of the nebular gas is computed by solving
Eq.~\ref{sigmaevol} with an explicit finite-difference technique on
a grid with a cell width $\Delta r \propto \sqrt{r}$. The resulting
type II migration forces on the giant planet are computed from
Eq.~\ref{adot} by deriving an instantaneous time scale, $a/\dot{a}$,
and inserting this into Eq.~6 of Paper I. Strong eccentricity
damping for the giant planet is assumed, with the damping time scale
being 1/50th of the radial migration timescale. As in Paper I, we
have neglected the effects of type I migration and associated
eccentricity damping \citep{ward2,tanaka2,tanaka3}.

The gas disk adopted here extends from an inner radius
$r_\mathrm{in} = 0.0025~\mathrm{AU}$ to an outer radius of
$r_\mathrm{out} = 33~\mathrm{AU}$ with an initial
$\Sigma_\mathrm{g}(r)$ profile given by Eq.~\ref{gas}. Since we are
considering a $3\times\mbox{MMSN}$ disk, $f_\mathrm{neb} = 3$ and
the initial disk mass is $M_\mathrm{gas} = 0.0398~\mathrm{M_\odot}
\approx 42~\mathrm{M_J}$. Note that this initial disk mass is
greater than the $2\times\mbox{MMSN}$ of gas assumed in Paper I;
however since that amount was kept fixed, the gas present in the
simulations presented here falls below this level after
$\sim$~140\,000 years. The boundary conditions for the
computation were $\Sigma_\mathrm{g}(r_\mathrm{in}) = 0$,
representing gas accretion onto the central star and at
$r_\mathrm{out}$ the radial velocity of the gas was set to zero.

To correctly couple the evolving gas disk algorithm with the N-body
code, synchronization of their respective time-steps is necessary.
In each simulation sub-run (see Sect.~\ref{initial}) the symplectic
N-body time-step $\tau_\mathrm{nbody}$ was fixed whereas the gas
disk time-step $\tau_\mathrm{gas}$ is adaptive and taken to be:

\begin{equation}\label{timestep}
\tau_\mathrm{gas} = \min \left( \frac{w~\Delta r(i)}{| v_r(i) |}
\right)\ ,
\end{equation}
where $i$ is the grid cell label and $w$ is a coefficient of order
unity that is tuned to ensure computational stability.
Including all the torques given above, the gas radial velocity is:

\begin{equation}\label{vr}
v_r = 2\sqrt{\frac{r}{GM_\ast}} \left( \Lambda +
\frac{T}{\Sigma_\mathrm{g} r} \right) - \frac{3}{\Sigma_\mathrm{g}
r^{1/2}} \frac{\partial}{\partial r} \left[\nu \Sigma_\mathrm{g}
r^{1/2} \right]\ .
\end{equation}
Thus, if $\tau_\mathrm{gas} \geqslant \tau_\mathrm{nbody}$ then we
set $\tau_\mathrm{gas} = \tau_\mathrm{nbody}$; if $\tau_\mathrm{gas}
< \tau_\mathrm{nbody}$ then the gas disk is evolved for
$\mathrm{INT}(\tau_\mathrm{nbody} / \tau_\mathrm{gas})$ steps of
duration $\tau_\mathrm{gas}$ plus an extra step of
$\mathrm{MOD}(\tau_\mathrm{nbody} / \tau_\mathrm{gas})$.

\subsection{Radial mixing of solid material.}\label{Radmixing}

At early times,
the solid component of a young protoplanetary nebula will exhibit a
radial pattern of chemical composition, controlled by the
temperature-dependent condensation sequence of a variety of metals,
rock minerals and ices. As the planetary system grows and evolves,
phenomena such as dynamical spreading, gas drag induced orbital
decay and resonant interactions can cause a radial mixing of
material. According to one school of thought
\citep[e.g.][]{morbidelli} the original matter that condensed in the
Earth's orbit is thought to have been dry \citep[for an alternative
opinion see][]{drake} and the origin of the Earth's water, and its
D/H ratio, can be explained if $\sim$~10\% of the planet's mass is
composed of carbonaceous chondrite-type material, originating from
between 2.5 -- 4~AU, and $\sim$~10\% of the water gained thereby is
retained at the end of accretion. N-body simulations of terrestrial
planet formation from disks that extend out close to the orbit of
Jupiter are supportive of this idea and all demonstrate substantial
mixing of water rich material into the inner disk
\citep[e.g.][]{morbidelli,chambers2,raymond1,raymond2,raymond3}.

In Paper I it was noted that one consequence of the inward migration
of a giant planet is the shepherding of planetesimals that are
damped by gas drag \citep{tanaka}, and the trapping of bodies at
first order mean motion resonances. Hence the outer, more
volatile-rich, regions of the protoplanetary disk are actively mixed
into its inner regions. However, the composition of planetesimals
and protoplanets, and their accretion products, were not logged in
our simulations.

We amend this here by labeling all particles with a composition
based on its original position in the disk, and summing the
composition of protoplanets as they grow. In a similar manner to
most other studies, we assume a crude three-phase initial radial
composition with rocky material originating at $<$~2~AU, material
characteristic of chondritic meteorites between 2 -- 2.7~AU, and
trans-snowline material at $>$~2.7~AU. For convenience, these phases
are referred to as ``dry", ``damp" and ``wet" respectively but, in
contrast to other studies, we do not assign an actual water mass
fraction to them. Instead we use our results to make more generalized
observations and predictions.

\subsection{Initial conditions and running of the simulations.}\label{initial}

\begin{table}
\caption{Data describing initial solids disk set-up} %
\label{table:1}  %
\centering
\begin{tabular}{c| c c| c}
 \hline\hline %
& Rocky Zone & Icy Zone& Total\\
& 0.4--2.7~AU & 2.7--4.0~AU & 0.4--4.0~AU\\
 \hline
$M_{\mathrm{solid}}$ & $9.99~\mathrm{M}_{\oplus}$ &
$14.8~\mathrm{M}_{\oplus}$ & $24.8~\mathrm{M}_{\oplus}$\\ %
 \hline
$m_{\mathrm{proto}}$ & $0.025~\mathrm{M}_{\oplus}$ &
$0.1~\mathrm{M}_{\oplus}$\\ %
$N$ & 66 & 9 & 75\\ %
 \hline
$m_{\mathrm{s-pl}}$ & $0.0025~\mathrm{M}_{\oplus}$ &
$0.01~\mathrm{M}_{\oplus}$\\ %
$N'$ & 3336 & 1392 & 4278\\
 \hline
$f_{\mathrm{proto}}$ & 0.17 & 0.06 & 0.1\\ %
 \hline\hline
\end{tabular}
\end{table}

\begin{figure}
 \resizebox{\hsize}{!}{\includegraphics{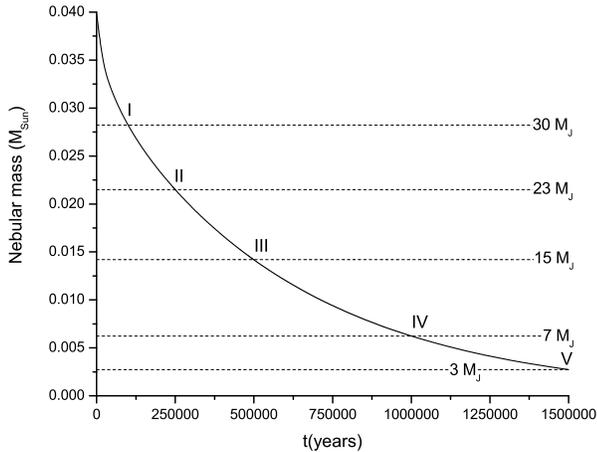}}
 \caption{Evolution of the mass of the gas disk. The mass of gas (in
  Jupiter masses) remaining at the launch point for each of the five
  migration scenarios is indicated.}
 \label{figure:1}
\end{figure}

\begin{figure}
 \resizebox{\hsize}{!}{\includegraphics{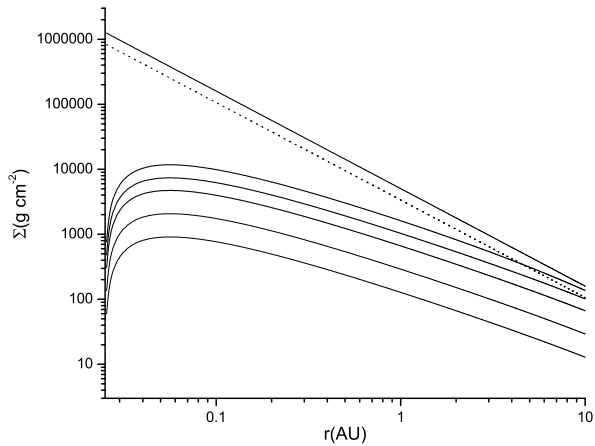}}
 \caption{Evolution of the gas surface density within the inner 10~AU
  of our simulated disk. The upper solid line is the
  $r^{-1.5}~\Sigma_\mathrm{g}$-profile for a $3\times\mbox{MMSN}$ disk. The
  lower solid curves, in descending order, are the profiles at 0.1,
  0.25, 0.5, 1.0 and 1.5~Myr respectively. The dashed line is the fixed
  $\Sigma_\mathrm{g}$-profile assumed in Paper I.}
 \label{figure:2}
\end{figure}

In common with Paper I, we assume a nominal age for the
protoplanetary disk of 0.5~Myr, this being the $t = 0$ start date
for our simulations, and a mass of three times that of the MMSN. Our
reasons for choosing a more massive nebula stem from the fact that
core-accretion theories of giant planet formation require an
enhanced density of solid material in order to grow a critical core
mass before the nebular gas is lost \citep{lissauer,thommes} which
might be supported by the observation that hot-Jupiters are found
predominantly around stars more metal-rich than the Sun
\citep{santos04a,fischer}. As runs proceed, the initial gas disk
described by Eq.~\ref{gas} evolves according to the algorithm
outlined in Sect.~\ref{gasmodel}, initially without the presence of
a giant planet (i.e. $\Lambda = 0, T = 0$ in Eq.~\ref{sigmaevol}).

Simultaneously, \emph{Mercury 6} evolves an initial disk of solid
material, extending radially from 0.4 -- 4.0~AU. We use the same
two-component initial solids disk as was used in Paper I, generated
according to the profile of Eq.~\ref{solids} and in line with the
oligarchic growth picture of \citet{kokubo}, which we assume to be a
reasonable description of the state of the inner disk at 0.5~Myr.
Data for this initial disk model are shown in Table~\ref{table:1}
which gives, for zones interior and exterior to the snowline, values
for the total mass of solid material $M_{\mathrm{solid}}$, the
number and mass of protoplanets $N$ and $m_{\mathrm{proto}}$, and
the number and mass of super-planetesimals $N'$ and
$m_{\mathrm{s-pl}}$. Note that as super-planetesimals act as
statistical tracers for a much more numerous population of real
planetesimals, their mass is much greater than that of a real
planetesimal, but for the purposes of calculating gas drag, each
super-planetesimal is treated as if it is a single 10~km radius
object of realistic mass. The parameter $f_\mathrm{proto}$, at the
foot of Table~\ref{table:1}, is the mass fraction of the solids disk
contained in protoplanets and we use this here as a rough measure of
the evolution of the disk, taking $f_\mathrm{proto} = 0.5$ to denote
the transition between oligarchic and giant impact growth regimes.

From $t = 0$, we run our new model for 0.1, 0.25, 0.5, 1.0 and
1.5~Myr in the absence of the giant, with
$\tau_\mathrm{nbody} = 8$~days and a simulation inner
edge of $r_* = 0.1~\mathrm{AU}$. (Note that $r_*$ does not
necessarily denote the physical radius of the central star, but it
is the distance interior to which particles are removed from the
simulation and their masses added to the central star.) The
evolution of the nebular mass over this time span and the particular
nebular mass at each of these five epochs are shown in
Fig.~\ref{figure:1}. The gas surface density profiles resulting at
these times are shown in Fig.~\ref{figure:2}. It is apparent that
gas drains onto the central star very rapidly at first, as the
density gradient relaxes to a shallower profile. Compared to the
unevolved profile, order of magnitude reductions in gas density
occur within the disk's inner regions. Data for the evolved solid
components are given in Table~\ref{table:2} and include the values
of $m_{\mathrm{max}}$, the mass of the largest protoplanet to have
evolved in each case. (In contrast to Paper I we do not run our
model to 3.0~Myr, as there remains too little gas at this time to
provide for our giant planet's envelope.) The advance of planetary
growth with time is indicated by the progressive increase of
$m_{\mathrm{max}}$ and $f_\mathrm{proto}$ and the reduction in
particle numbers as super-planetesimals are accreted and
protoplanets merge. When compared to the equivalent data from Paper
I, we note that no planetesimals are lost beyond the inner edge of
the simulation because of much lower gas densities near the central
star.

\begin{table}
\caption{Overall solids disk data: after 0.1--1.5 Myr of
evolution} %
\label{table:2}  %
\centering  %
\begin{tabular}{c| c c c c c}
 \hline\hline
Time~(Myr) & 0.1 & 0.25 & 0.5 & 1.0 & 1.5\\
Scenario ID & I & II & III & IV & V\\
 \hline
$M_{\mathrm{solid}}~(\mathrm{M}_{\oplus})$ & 24.8 & 24.8 & 24.8 & 24.8 & 24.8\\
$m_{\mathrm{max}}~(\mathrm{M}_{\oplus})$ & 0.29 & 0.37 & 0.63 & 1.40 & 2.13\\
$N$ & 50 & 47 & 44 & 38 & 34\\
$N'$ & 4031 & 3665 & 3036 & 2342 & 1964\\
$f_{\mathrm{proto}}$ & 0.19 & 0.26 & 0.37 & 0.51 & 0.59\\ %
 \hline\hline
\end{tabular}
\end{table}

\begin{figure}
 \resizebox{\hsize}{!}{\includegraphics{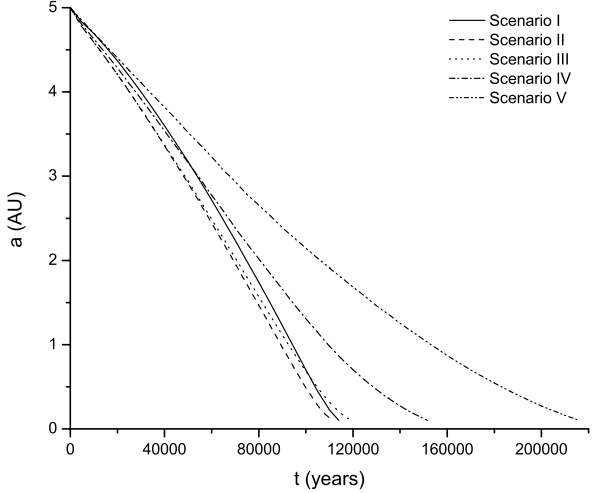}}
 \caption{Semi-major axis evolution of the giant planet in each
 scenario from the launch time (the top row in Table~\ref{table:2})
  to the time at which it arrives at 0.1~AU}
 \label{figure:3}
\end{figure}

The five type II migration scenarios studied here are constructed
from the five evolved disks indicated in Fig.~\ref{figure:1} and
summarized in Table~\ref{table:2}. Scenarios I -- III take place
whilst the solids disk remains in its oligarchic growth phase
($f_\mathrm{proto} < 0.5$) whereas Scenarios IV -- V have just
entered the final giant impact stage of growth ($f_\mathrm{proto} >
0.5$). A giant planet of mass $0.5~M_\mathrm{J}$ is placed into each
simulation at 5.0~AU after removing $0.4~M_\mathrm{J}$ of gas from
between 3 -- 7~AU. The giant then proceeds to clear out an annular
gap in the disk and migrates inward according to the method
described in Sect.~\ref{gasmodel}. The runs are halted once the
giant reaches 0.1~AU (see Fig.~\ref{figure:3}). For Scenarios I --
III, this takes $\sim 110\,000 - 120\,000$~years, close to the
prediction of Eq.~\ref{viscoustime}. The process takes longer to
complete in the cases of Scenarios IV -- V ($\sim$~150\,000 and
220\,000 years respectively) because by the time of the appearance
of the giant planet the gas disk is substantially depleted and is
less effective at driving migration. In order to better model
processes when the giant migrates down to small radial distances, we
contract the simulation inner edge down to a realistic T-Tauri star
radius: $r_* = 0.014~\mathrm{AU} \cong 3~R_\odot$. The
initial timestep chosen for the symplectic integrator was
$\tau_\mathrm{nbody} = 8$~days, but it was necessary to reduce this
at late times as material is driven into closer orbits. Hence each
scenario was divided into a number of sequential sub-runs with
$\tau_\mathrm{nbody}$ being adjusted at each re-start so as to keep
the timestep close to one tenth the orbital period of the innermost
object. Since planetesimals in this new model suffer less orbital
decay due to gas drag, it was possible to conduct these runs with a
higher value of $\tau_\mathrm{nbody}$ than in Paper I. However, this
advantage was negated at late times as the adaptive
$\tau_\mathrm{gas}$ falls steeply as the giant planet moves within
1~AU. From $t = 0$, the scenarios presented here each required
4 -- 8 weeks of 2.8~GHz CPU-time for completion.

\begin{figure*}
\centering
 \includegraphics[width=17cm]{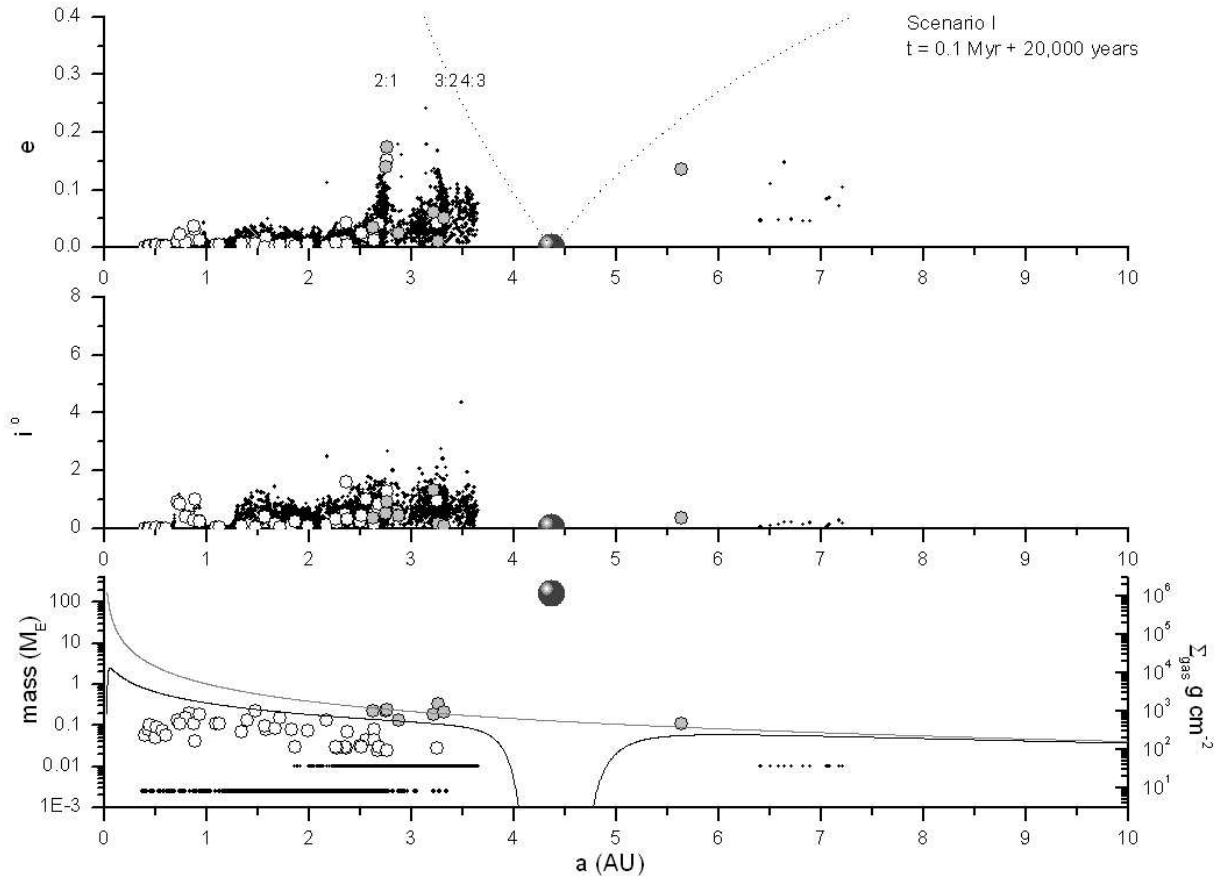}
 \caption{Scenario I at $20\,000$ years after the start of giant planet
 migration, showing the mass, inclination and eccentricity of objects.
 Small black dots represent super-planetesimals; white
 filled circles are rocky protoplanets; grey filled circles are
 icy protoplanets and the large highlighted grey filled circle is
 the giant. Objects plotted between the dotted lines in the upper panel
 have orbits that intersect the orbit of the giant. The location of the
 2:1, 3:2 and 4:3 resonances with the giant are indicated. Gas surface
 density is read on the right hand axis of the lower panel, the upper
 grey curve being the unevolved profile at $t$ = 0 and the lower black
 curve being the current profile.}
 \label{figure:4}
\end{figure*}

\section{Results.}\label{results}

We begin describing the results by focussing on the evolution during
the migration of the giant planet, discussing one specific case in
detail before examining differences between the various runs. We
then go onto describe the evolution of Scenario V after giant planet
migration has halted, focussing on the issue of terrestrial planet
formation in the scattered disk, and the likely composition of
planets that form there.

\subsection{Evolution during giant planet migration}
\label{migration-results}

The results of all scenarios showed a number of behavioral features
in common. As in Paper I, we illustrate these first by describing
the results of Scenario I in detail. We then proceed to examine how
the results differ between scenarios (dependence on disk maturity)
and how the results differ from those of Paper I (dependence on an
evolving gas disk).

\subsubsection{Typical features of a run.}\label{features}

The typical effects of a migrating giant planet on an inner solids
disk observed from Paper I were as follows: 1) shepherding of
planetesimals; 2) capture of objects at first order mean motion
resonances; 3) acceleration of accretion interior to the giant with
possible hot-Neptune formation; and 4) generation of a scattered
exterior disk. To a greater or lesser extent, these features were
also observed in our new simulations. Four snapshots of the
evolution of Scenario I are illustrated in Figs.~4 -- 7 showing the
mass, inclination and eccentricity of objects, and the gas surface
density vs. semi-major axis. The original provenance of the
protoplanets (interior or exterior to the snowline) is denoted by
the shading of its symbol as described in the caption to
Fig.~\ref{figure:4}. In the case of a merger between rocky and icy
protoplanets, this shading is determined by that of the most massive
of the pair.

\begin{figure*}
\centering
 \includegraphics[width=17cm]{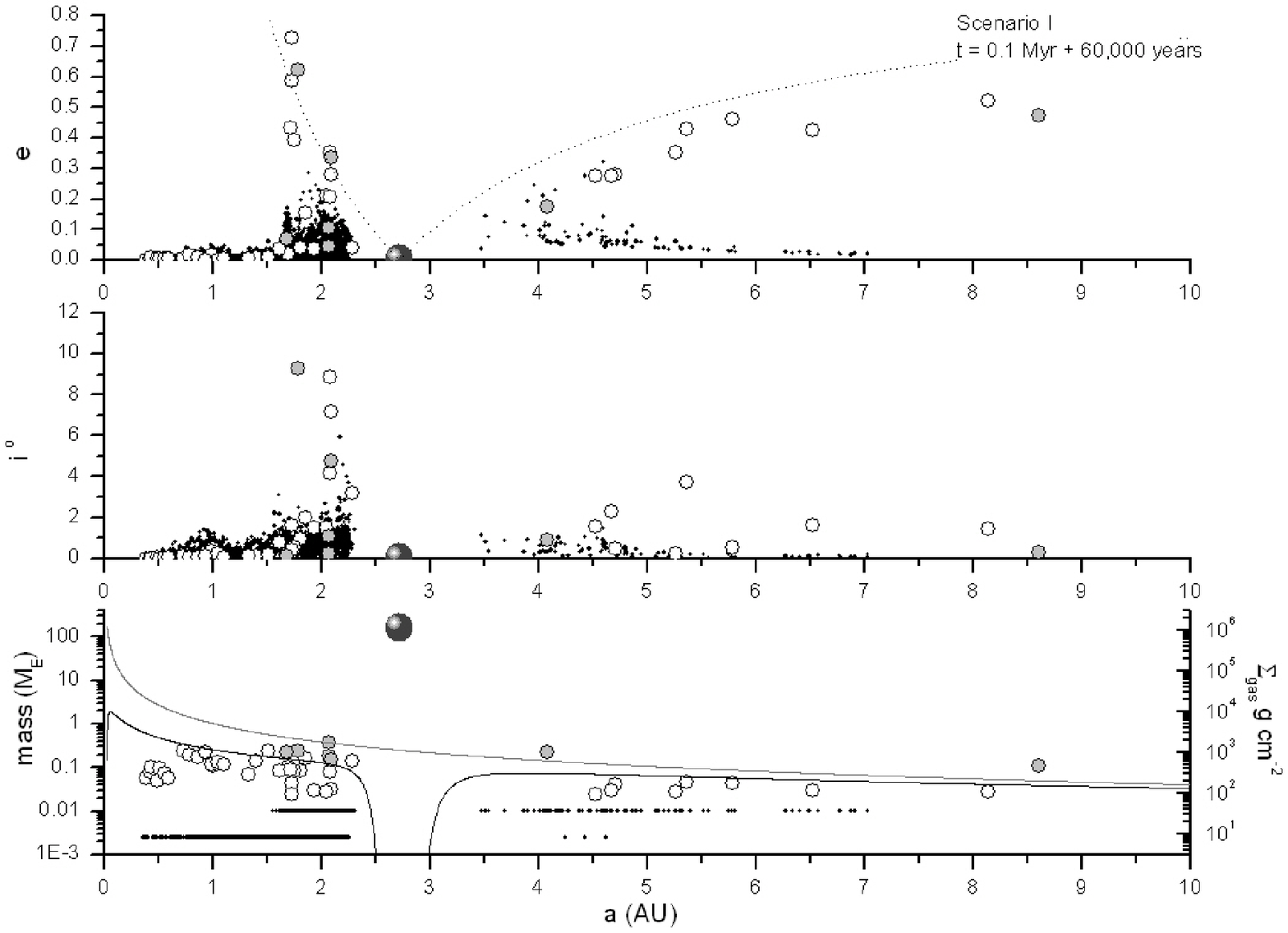}
 \caption{Scenario I at $60\,000$ years after the start of giant planet
 migration. The giant has now moved inward to 2.72 AU. Increasing
 excitation of the orbits of protoplanets captured at resonances is
 apparent, as is the build-up of matter scattered into external orbits.}
 \label{figure:5}
\end{figure*}

\begin{figure*}
\centering
 \includegraphics[width=17cm]{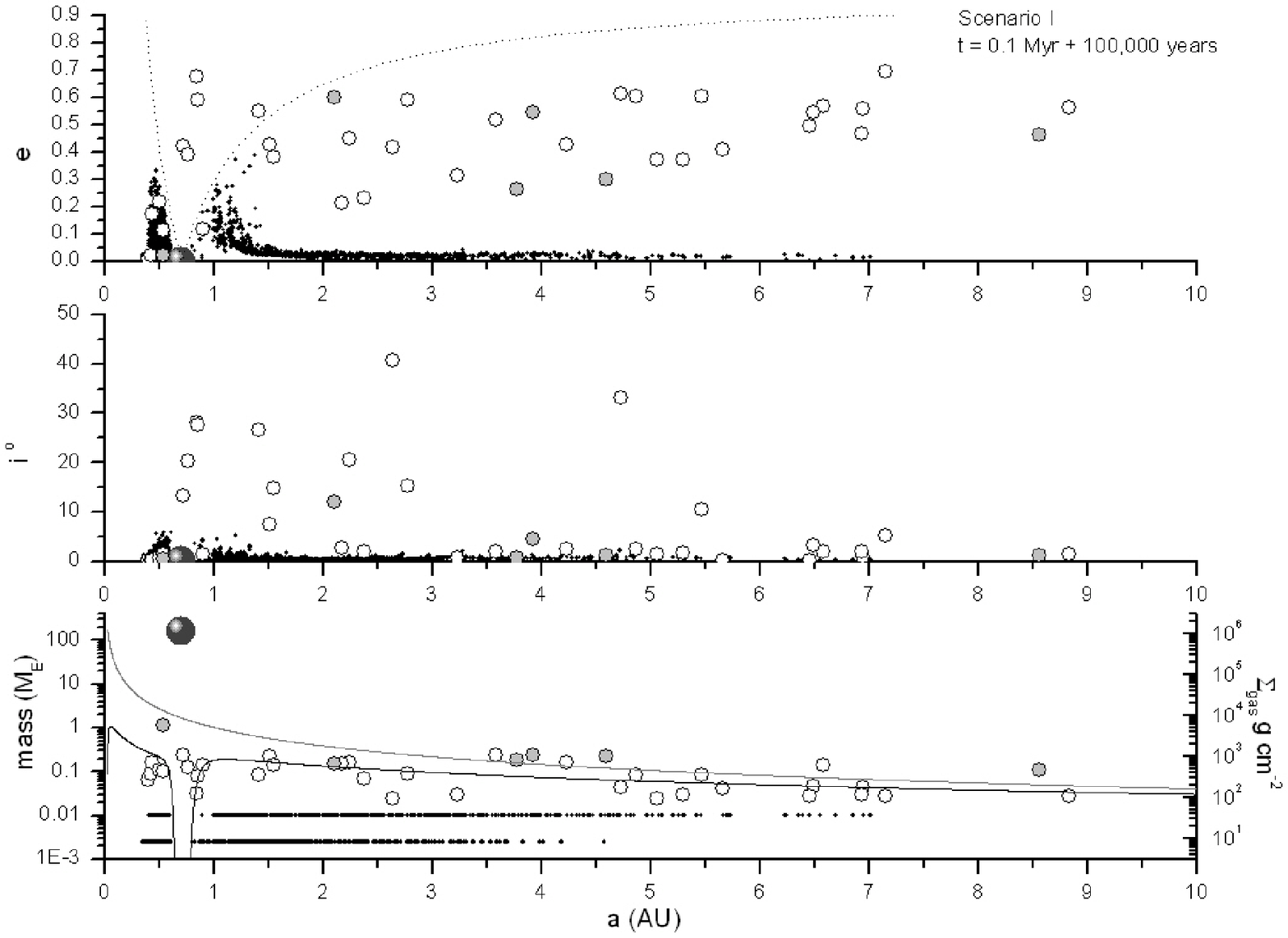}
 \caption{Scenario I at $100\,000$ years after the start of giant planet
 migration. The giant has now moved inward to 0.70 AU. Five protoplanets
 are currently crossing the orbit of the giant. The scattered disk has
 grown and a $> 1~\mathrm{m_\oplus}$ planet is accreting within the
 compacted interior disk.}
 \label{figure:6}
\end{figure*}

An early stage in the evolution of Scenario I, 20\,000 years after
the introduction of the giant planet, is shown in
Fig.~\ref{figure:4}. The giant has opened a $\sim 0.75$~AU gap in
the gas and has migrated inward to 4.37~AU, shepherding the outer
disk edge at the 4:3 resonance, now at 3.61~AU. Capture of objects
at the 3:2 and 2:1 resonances, at 3.33 and 2.75~AU respectively, is
apparent from eccentricity spikes visible in the upper panel and a
clustering of protoplanets in the lower panel. Even at this early
phase, before the giant has entered the original confines of the
interior disk, one protoplanet and a handful of super-planetesimals
have been scattered into external orbits.

The system midway through the run, 60\,000 years after the
introduction of the giant planet, is shown in Fig.~\ref{figure:5}.
The giant has now migrated to 2.72~AU, putting the positions of the
3:2 and 2:1 resonances at 2.07 and 1.71~AU respectively. Strong
excitation of protoplanetary orbits is now apparent at these
locations, as is the build-up of scattered material in external
orbits. The primary mechanism of this expulsion is evident from the
behavior of material captured at resonances. Continuous resonant
pumping results in orbits becoming eccentric enough to eventually
intersect the orbit of the giant planet. A series of close
encounters with the giant then follows, eventually resulting in a
final encounter where the object is accreted or expelled into a
non-intersecting exterior orbit.

\begin{figure*}
\centering
 \includegraphics[width=17cm]{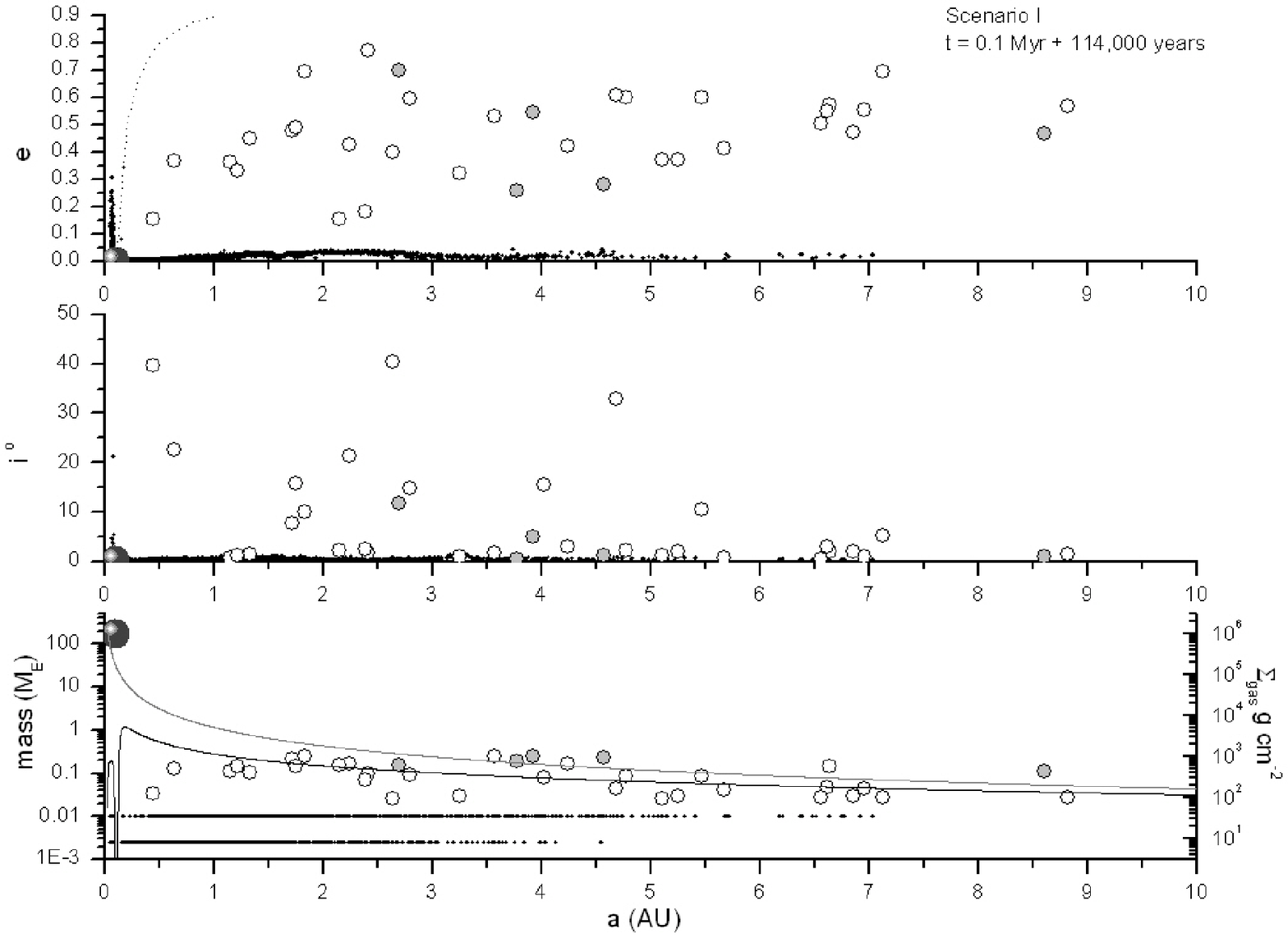}
 \caption{Scenario I at $114\,000$ years after the start of giant planet
 migration. The giant planet has migrated to 0.1 AU. Most interior mass has
 been lost after the most massive interior protoplanet impacts the giant.
 63\% of the original solids disk mass now resides in exterior orbits.}
 \label{figure:7}
\end{figure*}

An advanced stage of Scenario I, 100\,000 years after the
introduction of the giant planet is shown in Fig.~\ref{figure:6}.
The giant planet is now at 0.70~AU and the 3:2 and 2:1 resonances
are at 0.54 and 0.44~AU respectively. A substantial scattered
external disk has now formed and sufficient gas remains in this
early scenario to rapidly damp the orbits of scattered
planetesimals. An impression of the scattering process in action is
given by the five protoplanets currently crossing the giant's orbit.
The interior disk is compacted to high surface densities, but now
that strong first order resonances with the giant are influential
throughout its width, and gas densities have fallen by a factor of
$>$~10, there is a noticeable dynamical stirring of its entire
remaining population. Nevertheless, accretion has speeded up in this
shepherded zone with the growth of one protoplanet of
1.17~$\mathrm{m_\oplus}$ at 0.53~AU outrunning that of its
neighbors.

\begin{figure*}
\centering
 \includegraphics[width=17cm]{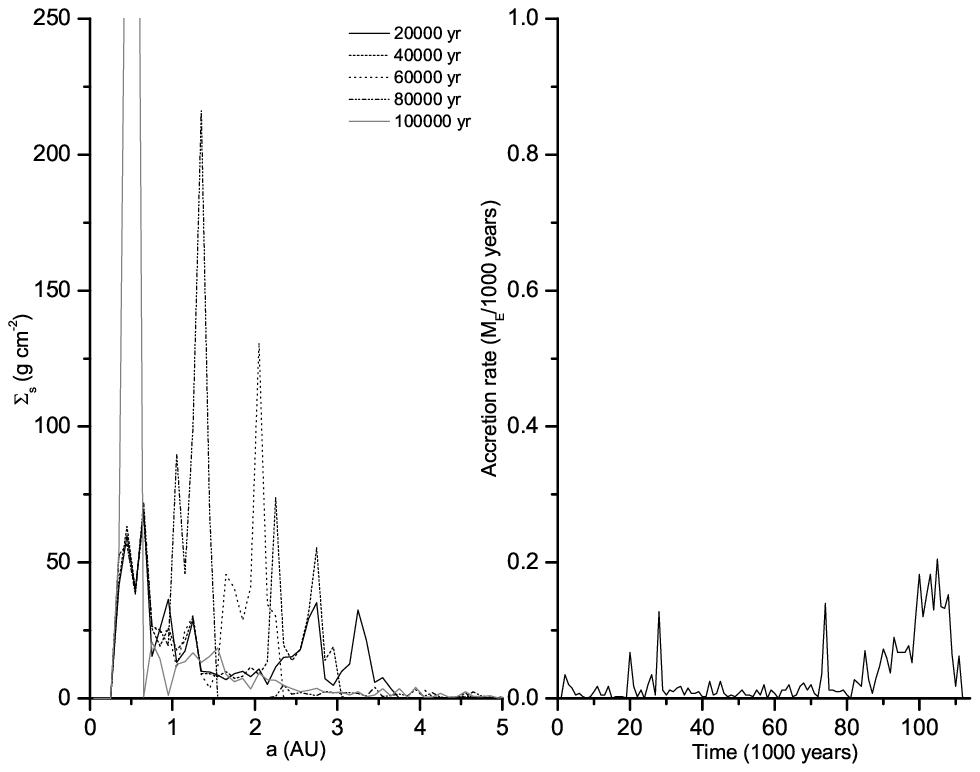}
 \caption{Surface density evolution (left hand panel) and accretion rates
 (right hand panel) for Scenario I. Growing surface density peaks at the 2:1
 and 3:2 resonances sweep through the inner system ahead of the giant.
 Accretion rates increase after $\sim 80\,000$ years within the compacted
 portion of the disk.}
 \label{figure:8}
\end{figure*}

Scenario I is terminated at 114\,000 years after the start of
migration when the giant planet arrives at 0.1~AU and the system is
illustrated at this point in Fig.~\ref{figure:7}. Two thirds of the
original solids disk mass has survived the migration episode -- the
great majority of this residing in the scattered exterior disk. Most
of the remainder has been accreted by the giant planet. Just $\sim
4\%$ of the original disk mass remains interior to the giant and
none of this is in the form of large bodies. The rapidly accreting
interior protoplanet seen in Fig.~\ref{figure:6} continued its
inward progress close to the 3:2 resonance position with its orbit
being well-damped by strong collisional damping and dynamical
friction from planetesimals and smaller protoplanets (see
Fig.~\ref{figure:10} in Sect.~\ref{inner-planets}). However, $\sim
2000$~years before its demise, the protoplanet drifted outward and
became captured at the 4:3 resonance. At this location, both
accretion and dynamical friction were reduced allowing the
protoplanet's orbit to become progressively more eccentric. At
109\,220 years the protoplanet, now weighing in at
2.41~$\mathrm{m_\oplus}$, collided with and was accreted by the
giant. The five other less massive interior planets visible in
Fig.~\ref{figure:6} grew very little, remaining between $\sim 0.03 -
0.2 \mathrm{m_\oplus}$, and in due course one of them impacted the
giant and four were scattered.

To emphasize the above description of Scenario I, the surface
density evolution of the disk and its accretion rate are shown in
Fig.~\ref{figure:8}. The left hand panel shows the disk surface
density profile (obtained by summing all protoplanets and
super-planetesimals in 0.1~AU width bins) at 20\,000, 40\,000,
60\,000, 80\,000 and 100\,000 years after the start of migration;
the right hand panel plots the amount of mass accreted onto
protoplanets only (including protoplanet mergers) every 1\,000 years
for the duration of the run. In the $\Sigma_{\mathrm{s}}$ plot, two
surface density enhancements are clearly visible as spikes at the
3:2 and 2:1 resonances and are seen to grow whilst moving inward. At
80\,000 years, these have almost merged into one: the shepherded
portion of the original disk having by now been squeezed into a
dense ring. By 100\,000 years, most of this mass is now confined
within 0.6 AU and $\Sigma_{\mathrm{s}}$ here has risen to
$\sim~500~\mathrm{g~cm}^{-2}$ which is off the vertical scale in the
figure. This amounts to an increase by a factor of $\sim$~10 over
the previous, undisturbed, disk surface density, but is only about
half the increase seen in the equivalent Scenario presented in Paper
I. The effect of this disk compaction process is visible in the
accretion rate plot. Mass accretion rises significantly after
80\,000 years due to both the high values of $\Sigma_{\mathrm{s}}$
and the fact that much of this mass now resides in a zone where
dynamical times are shorter. However, the large, terminal, accretion
rate spike described in Paper I is not reproduced here (compare
Fig.~\ref{figure:8} with Fig.~6 of Paper I). This is because close
to the end of that previous simulation a 15.65~$\mathrm{m_\oplus}$
hot-Neptune was assembled in a dramatic phase of runaway accretion
interior to 0.1~AU. In the case presented here the compacted
interior disk is only half as dense and is much less well damped
(note the large difference between the upper and lower gas density
curves in Fig.~\ref{figure:7}) and, whilst a protoplanet does grow
to 2.41~$\mathrm{m_\oplus}$ in this region, as described above, it
does not survive and is accreted by the giant planet.
We describe the formation and fate of interior planets
in more detail in Sect.~\ref{inner-planets}.

\subsubsection{Dependence on the maturity of the inner
disk.}\label{maturity}

The reason for running five scenarios through a progressively more
mature inner disk is to see if the timing of migration has any
systematic effect on the results. This is possible as the partitioning
of the solids disk between inner and outer remnants is influenced by
the level of damping that particles are subject to, which declines
with age. In Paper I, where $\Sigma_g(r)$ is fixed, this occurs as a
side effect of accretion: as planetesimals are accreted by
protoplanets ($f_\mathrm{proto}$ increases), fewer small particles
remain that are subject to gas drag and which can exert dynamical
friction. In these latest simulations, since we now have an evolving
gas disk, the strength of gas drag on susceptible particles also
declines with time and is particularly marked close to the central
star and giant planet.

\begin{table*}
\caption{Fate of the disk mass at the end of each run.} %
\label{table:3}  %
\centering  %
\begin{tabular}{c| c c c c c}
 \hline\hline %
Scenario & I & II & III & IV & V\\
 \hline\hline %
Total Initial Solids $(\mathrm{M}_{\oplus})$ & 24.81 & 24.81 & 24.81
& 24.81 & 24.81\\
 \hline %
Total Surviving Solids $(\mathrm{M}_{\oplus})$ & 16.60~(67\%) &
16.69~(67\%) & 17.40~(70\%) & 21.23~(86\%) & 20.22~(81\%)\\
 \hline %
Interior Surviving Solids $(\mathrm{M}_{\oplus})$ & 0.88~(4\%) &
0.65~(2\%) & 1.00~(4\%) & 0.84~(3\%) & 0.31~(1\%)\\
$N, f_{\mathrm{proto}}$ & 0, 0 & 0, 0 & 0, 0 & 0, 0 & 0, 0\\
 \hline %
Exterior Surviving Solids $(\mathrm{M}_{\oplus})$ & 15.72~(63\%) &
16.04~(65\%) & 16.40~(66\%) & 20.39~(82\%) & 19.90~(80\%)\\
$N, f_{\mathrm{proto}}$ & 39, 0.27 & 29, 0.28 & 33, 0.42 & 31, 0.63
& 23, 0.66\\
 \hline %
Accreted by Star $(\mathrm{M}_{\oplus})$ & 0.01~(0.04\%) &
0.01~(0.04\%) &
0.11~(0.4\%) & 0.0~(0\%) & 0.0~(0\%)\\
 \hline %
Accreted by Giant $(\mathrm{M}_{\oplus})$ & 8.20~(33\%) &
7.85~(32\%)
& 6.77~(27\%) & 3.41~(14\%) & 4.59~(19\%)\\
 \hline %
Ejected $(\mathrm{M}_{\oplus})$ & 0.00~(0\%) & 0.26~(1\%) &
0.51~(2\%) & 0.17~(1\%) & 0.0~(0\%)\\
 \hline\hline %
\end{tabular}
\end{table*}

Data describing the fate of the solids disk mass at the end of each
scenario are shown in Table \ref{table:3}. Disk mass that is lost is
either accreted by the central star, ejected from the system, or
accreted by the giant planet; that which survives is partitioned
between bodies orbiting interior or exterior to the final orbit of
the giant planet at 0.1~AU.

In all scenarios, a negligible quantity of mass was ejected or lost
to the central star. However, a significant fraction of the disk
mass (14 -- 33\%) was accreted by the giant, especially towards the
end of the migration. At these late times, planetesimals are
shepherded into the partially evacuated inner regions of the gas
disk where gas drag is less effective at damping orbital
perturbations from the giant and growing protoplanets. Once a
planetesimal strays into the annular gap in the gas containing the
giant, gas drag vanishes and accretion or scattering by the giant
follows. This increased excitation of the shepherded planetesimal
population, and the thinning down of their number, renders them less
effective at damping protoplanetary orbits via dynamical friction
and collisions. Hence, at late times the orbits of the remaining
interior protoplanets also tend to destabilize, with one of the same
two fates in store. A trend can be seen in Table \ref{table:3} for
the giant planet to accrete less material with disk maturity. This
occurs because as the disk ages the gas density and the solids mass
fraction in small bodies both decline, resulting in less dynamical
dissipation of both planetesimals and protoplanets. Less matter is
shepherded in such mature disks so there is less of an interior
remnant for the giant to accrete from at late times.

In all scenarios, a large majority of the disk solids are found to
survive the migration episode -- over two thirds of the original
inventory. Table \ref{table:3} shows that there is essentially no
trend with disk maturity in the partitioning of surviving mass
between interior and exterior remnants. Just a few percent of the
mass remains interior to the giant in all cases. When the giant
planet migrates through a disk undergoing oligarchic growth
($f_\mathrm{proto} < 0.5$; Scenarios I -- III) $\sim$ 65\% of the
original disk mass survives by being scattered into the exterior
disk. This fraction increases for disks undergoing giant
impact-style growth ($f_\mathrm{proto} > 0.5$; Scenarios IV -- V) to
$>$ 80\%, not because more mass remains at $<$ 0.1~AU but because
less mass is accreted by the giant at late times.

\subsubsection{Dependence on an evolving gas disk.}\label{gasdependence}

The salient dynamical behaviors of solids disk particles such as
shepherding, resonant capture, scattering by the giant planet and
eventual partition into interior and exterior remnants are observed
generally in the results of both the present model and those of
Paper I. However, the introduction of an evolving gas disk causes
the relative predominance of these outcomes to differ. This is
because both the principal sources of dissipation, dynamical
friction and gas drag, fall with time, whereas in the model from
Paper I only the former declines. There are five systematic
differences between the results of Paper I and those presented here.

\begin{enumerate}
\item Much less mass is lost to the central star.
\item Much more mass is accreted by the giant planet.
\item Disk partitioning between inner and outer remnants
is much less sensitive to disk maturity.
\item The protoplanet mass fraction
($f_\mathrm{proto}$) in the exterior scattered disk is lower.
\item Interior hot-Neptune-type planets grow to smaller masses
and do not survive.
\end{enumerate}

The reasons for the first two items are the large reductions in gas
density close to the central star and in the vicinity of the giant
planet. Planetesimals do not have time to spiral into the central
star and feel little eccentricity damping when close to the giant because
of the low gas drag in these regions. The reasons for the last three
items stem from the fact that the system is less dissipative so
scattering behavior predominates over shepherding at all epochs.
Interior remnants are consistently much smaller than outer remnants
which always include $>$ 60\% of the original solids disk mass. The
increased tendency for planetesimals to scatter means that there is
less of a fractionation of planetesimals from protoplanets between
interior and exterior remnants. More planetesimals are expelled into
the exterior disk and hence its overall mass fraction contained in
protoplanets ($f_\mathrm{proto}$) is lower.

\begin{figure}
 \resizebox{\hsize}{!}{\includegraphics{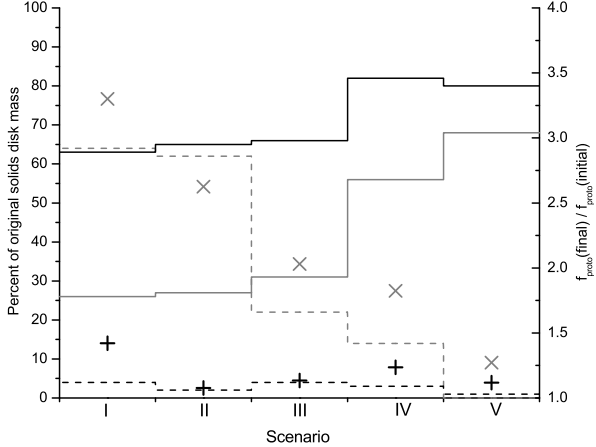}}
 \caption{Comparison of the results of Paper I (grey lines and $\times$~symbols)
  and the results of this paper (black lines and $+$~symbols). Dashed lines and
  solid lines are the percentage of original disk mass found in the
  interior and exterior remnants respectively. The ratios of
  the protoplanet mass fraction in the final exterior disk to that of
  the original disk, $f_\mathrm{proto}(\mathrm{final}) \ /
  f_\mathrm{proto}(\mathrm{initial})$, are indicated
  by the $\times$ and $+$ symbols and are read off the right hand y-axis.}
 \label{figure:9}
\end{figure}

These tendencies are illustrated in Fig.~\ref{figure:9} where the
percent of the original solids disk mass surviving at the end of the
simulations presented here are compared with the
simulation results from Paper I.
Interior and exterior remnants are plotted as dashed and solid lines
respectively; gray lines are the results of Paper I and black lines
are those of the present work. It can be seen that partitioning
varies strongly with disk maturity in the case of a steady-state gas
model and weakly, if at all, when the nebula gas is allowed to
viscously evolve. Since gas drag does not decline in the former
case, the shepherding of planetesimals is more influential and more
mass remains in the interior remnant in early scenarios, which is
mostly contained in surviving hot-Neptune-type planets. Late
scenarios from Paper I behave more similarly to the ones presented
here as a greater fraction of the solids disk mass is contained
within protoplanets, which do not feel the gas, and their source of
dissipation, dynamical friction, is weaker due to a decline in
super-planetesimal numbers. \footnote{Note that Scenario V is less
comparable to its counterpart in Paper I as it is initiated at a
different time, 1.5~Myr earlier.}

The increased tendency in the present model for planetesimals to
scatter into the exterior disk can also be appreciated from
Fig.~\ref{figure:9} by comparing, for the two models, the ratio of
the protoplanet mass fraction in the final exterior disk to that of
the original disk, $f_\mathrm{proto}(\mathrm{final}) /
f_\mathrm{proto}(\mathrm{initial})$. If the scattering process does
not discriminate between protoplanets and planetesimals, and there
is no further accretion after scattering, this ratio would remain at
$f_\mathrm{proto}(\mathrm{final}) /
f_\mathrm{proto}(\mathrm{initial}) = 1$. Subsequent accretion will
raise this ratio, but not by much over the timescale considered,
especially as accretion rates are reduced in the exterior disk due
to its dynamically excited state. Reading from the right hand
y-axis, Fig.~\ref{figure:9} shows that
$f_\mathrm{proto}(\mathrm{final}) /
f_\mathrm{proto}(\mathrm{initial}) < 1.5$ for all of the present
model results whereas much higher values are obtained from Paper I.
Protoplanets are preferentially scattered into the exterior disk
when there is strong and unvarying gas drag, whereas fractionation
is much less marked when gas drag declines over time and falls to
near zero in the vicinity of the giant planet.

\subsubsection{Formation and survival of hot--Neptunes}
\label{inner-planets} 
\begin{figure}
 \resizebox{\hsize}{!}{\includegraphics{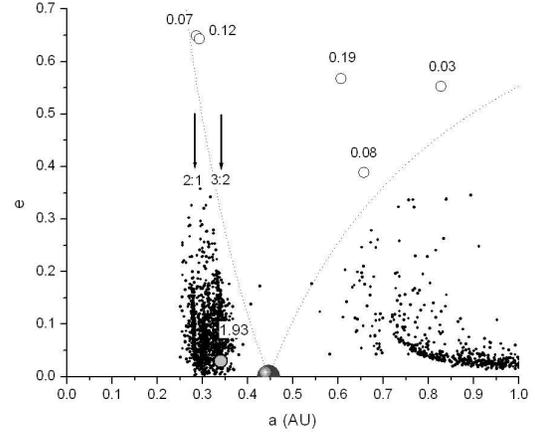}}
 \caption{Blow-up of the inner 1~AU of Scenario I, showing eccentricity
 vs semi-major axis 105\,000 years after
 the start of giant planet migration. Protoplanetary masses are indicated in
 $\mathrm{m}_\oplus$ and the locations of the 2:1 and 3:2 resonances are
 shown with arrows. The giant planet is at 0.48~AU and
 our hot-Earth candidate is a $1.93~\mathrm{m}_\oplus$ planet at 0.34~AU
 with $e \approx 0.03$. Five lower mass protoplanets are in the process of
 being scattered into the exterior disk. (Compare with Fig.~\ref{figure:6}
 which shows the situation 5\,000 years earlier.) As described in
 Sect.~\ref{features}, the hot-Earth candidate is eventually accreted
 by the giant planet $\sim$~4\,000 years later once it has migrated further
 inward to $\sim$~0.26~AU.}
 \label{figure:10}
\end{figure}

A striking feature of the results of Paper I
was the growth and survival, interior to the final orbit of the
giant, of hot-Neptune and hot-Earth type planets ranging between
$\sim 2 - 16~\mathrm{m}_\oplus$. No such planets are found to
survive in the runs of our present model. Accelerated accretion in
the shepherded disk is observed with some protoplanets growing to
$\sim 1 - 3~\mathrm{m}_\oplus$ (see Figs.~\ref{figure:6} and
\ref{figure:8}), but in each scenario their eventual fate was to
impact the giant planet just before the end of the simulation. After
formation, these planets become locked in a mean motion resonance
(3:2 or 4:3) with the giant, and subsequent eccentricity pumping
during migration leads to collision and merger with the giant
shortly before migration halts. However, had we halted giant planet
migration earlier, at $\sim$ 0.25 -- 0.5~AU instead of 0.1 AU, these
simulations would predict the existence of hot--Earths \textbf{(e.g.
see Fig.~\ref{figure:10})}.

We have performed additional calculations, very similar to those
presented here, in which we switched off the effect of dissipating
spiral waves travelling into the disk (i.e. $T(r)=0$ in
Eq.~\ref{sigmaevol}). This term helps create an inner cavity in the
gas disk as the planet migrates inward. These calculations produced
interior planets ranging in mass $\sim 2$ -- 4 $m_{\oplus}$ during
giant planet migration. A variety of final outcomes were noted: 1)
some of them were accreted by the giant planet; 2) others were
scattered externally into stable orbits at $a \sim 0.4$~AU; and 3) a
few remained in interior orbits, typically close to 0.076~AU (the
3:2 resonance). These results, taken alongside the five scenarios
already presented in this paper and those from Paper I show a
distinct trend: a strongly dissipating gas disk interior to the
giant planet leads to the formation of fairly massive,
hot--Neptune--like planets which survive; a gas disk of lower
density leads to the formation of lower mass interior planets that
often do not survive.

We note at this point that our models currently neglect some
potentially important sources of dissipation due to general
uncertainty about how planetary formation can proceed in their
presence, such as type I migration \citep{ward2,tanaka2,tanaka3} and
the circularization of orbits due to stellar tides. Associated with
the former is strong eccentricity and inclination damping of low
mass planets which may facilitate the survival of inner planets.
Simulations are currently underway to examine this possibility.
Tidal damping of orbits however is unlikely to have a significant
effect on our results because all shepherded protoplanets are
scattered or accreted by the giant at $a \gtrsim 0.1$~AU. Whilst
tidal circularization times at these radii are uncertain, due to the
uncertainty of $Q$, they probably range from $\sim 10^8 - 10^{10}$
yr \citep{goldreich}, orders of magnitude greater than the millenia
it takes for our giant planets to traverse the final 0.5~AU of their
migration.

\begin{table*}
\caption{Data describing the external surviving protoplanets at the end
of giant planet migration.} %
\label{table:4}  %
\centering  %
\begin{tabular}{c| c c c c c c c c c c c}
 \hline\hline %
Scenario & N & $\overline{m}_{\mathrm{proto}}~(\mathrm{M}_{\oplus})$
& $m_{\mathrm{max}}~(\mathrm{M}_{\oplus})$ &
$\overline{a}~(\mathrm{AU})$ & $a_{\mathrm{min}}~(\mathrm{AU})$ &
$a_{\mathrm{max}}~(\mathrm{AU})$ & $\overline{e}$ &
$e_{\mathrm{min}}$ & $e_{\mathrm{max}}$ &
$\overline{i}~\degr$ & $i_{\mathrm{max}}~\degr$\\
 \hline
I & 39 & 0.11 & 0.25 & 5.43 & 0.45 & 37.92 & 0.51 & 0.15 & 0.99 &
8.85 & 40.45\\
II & 29 & 0.16 & 0.74 & 4.99 & 0.71 & 15.25 & 0.49 & 0.045 & 0.91 &
8.59 & 35.45\\
III & 33 & 0.21 & 0.64 & 4.68 & 0.37 & 12.22 & 0.46 & 0.059 & 0.89
& 8.32 & 43.97\\
IV & 31 & 0.41 & 2.35 & 5.16 & 0.43 & 8.92 & 0.50 & 0.086 & 0.94 &
5.81 & 41.93\\
V & 23 & 0.57 & 2.19 & 6.88 & 0.49 & 18.18 & 0.45 & 0.079 & 0.83 &
3.53 & 16.24\\
 \hline\hline
\end{tabular}
\end{table*}

\subsubsection{The exterior scattered disk.}\label{exterior}

In Paper I it was noted that the exterior solids disks generated by
the giant planet migration were dynamically excited,
depleted of planetesimals, and
spread over a greater radial extent than the original disk. Whilst
subsequent planet formation still seemed possible in such a disk, it
was predicted to occur over significantly extended timescales.
However, in
the Paper I scenarios only $\sim 1 - 2 \mathrm{m_\oplus}$ of
planetesimals were scattered into the exterior disk, whereas in the
present model this quantity rises to $\sim 8 - 11
\mathrm{m_\oplus}$. In addition, planetesimals are not scattered as
widely as protoplanets and their excited orbits damp rapidly when
remote from the influence of the giant planet. Whilst the decline in
gas drag with time lessens this trend in later scenarios, in all
cases the inner regions of the scattered disk remain well populated.

There is a greater similarity between the current models in the outcome of
scattering of the protoplanetary population and data for the
external protoplanets are shown in Table \ref{table:4}, giving their
number, mean and maximum masses and orbital inclinations, and their
mean, minimum and maximum semi major axes and eccentricities. As
expected, protoplanetary numbers fall and masses rise with disk
maturity, an effect largely due to prior accretion before the
appearance of the giant planet. Allowing for stochastic events, such
as giant impacts and strong scatterings, these data are much the
same in Paper I, as are the mean semi-major axes and eccentricities
($\bar{a} \approx 5~\mathrm{AU}, \bar{e} \approx 0.5$). However the
minimum values of the semi-major axes and eccentricities
($a_\mathrm{min}$ and $e_\mathrm{min}$) are lower than in Paper I. This
is because late-shepherded protoplanets tend to scatter rather than
assembling into hot-Neptunes and their resultant orbits damp more
rapidly as many more planetesimals are available to exert dynamical
friction.

\subsection{Post--migration terrestrial planet formation}
\label{outer-planets}

\begin{figure}
 \resizebox{\hsize}{!}{\includegraphics{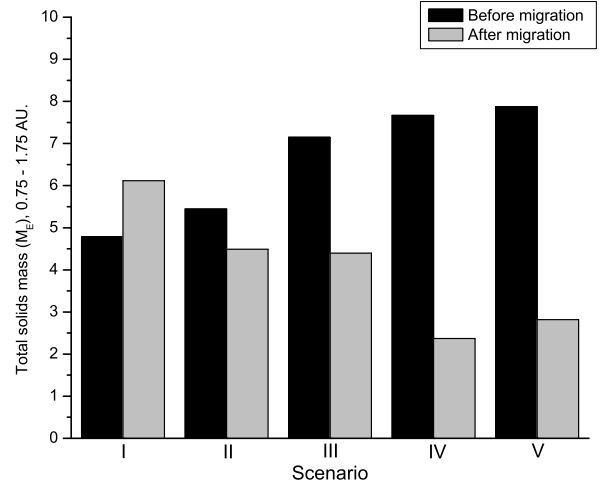}}
 \caption{The total solids mass between 0.75 -- 1.75~AU,
 before and after the giant planet
 migration, plotted for each scenario.}
 \label{figure:11}
\end{figure}

What do the results of the five scenarios presented here have to say
about the probablity of forming terrestrial planets in the scattered
disks ? Considering the Scenarios I -- V overall, if the factor of
long--term importance is the mass distribution, rather than initial
dynamics, then such planets should form and await discovery in
hot--Jupiter systems. This point is made in Fig.~\ref{figure:11}
where the total solids mass with semi-major axes between 0.75 --
1.75~AU, before and after giant planet migration, is plotted for
each scenario. A clear trend is visible for less matter to be found
in this region with increasing disk maturity as it is more widely
scattered. Mass dispersal in late scenarios, however, is partially
offset by the pre-existing inward evolution of material in more
mature disks which enhances the mass present in inner regions. It is
also offset by the fact that fairly massive protoplanets are
scattered into the external disk in late Scenarios (IV and V -- see
table~\ref{table:4}), so that terrestrial planet formation in the
scattered disk has already received a significant boost from
accretion prior to and during migration. In all cases, more than
$2~\mathrm{m}_\oplus$ of planet forming material remains in the
`maximum greenhouse' habitable zone  \citep{kasting} of the system
after the passage of the giant planet.

Extending our runs for the additional 50 -- 100 Myr it would take to
form a completed external planetary system is beyond our current
computational capabilities. The principal difficulty is the presence
of the hot--Jupiter at 0.1 AU which limits $\tau_\mathrm{nbody}$ to
an excessively low value. We have, however, extended Scenario V for
2~Myr and find that (apparently stable) terrestrial planets {\emph
do} form in the habitable zone of this system.

As the gas density is low when migration of the giant planet halts
in Scenario V, we assume rapid removal of the remnant gas {\it via}
photoevaporation and evolve the system in the absence of gas.
\begin{figure*}
\centering
 \includegraphics[width=17cm]{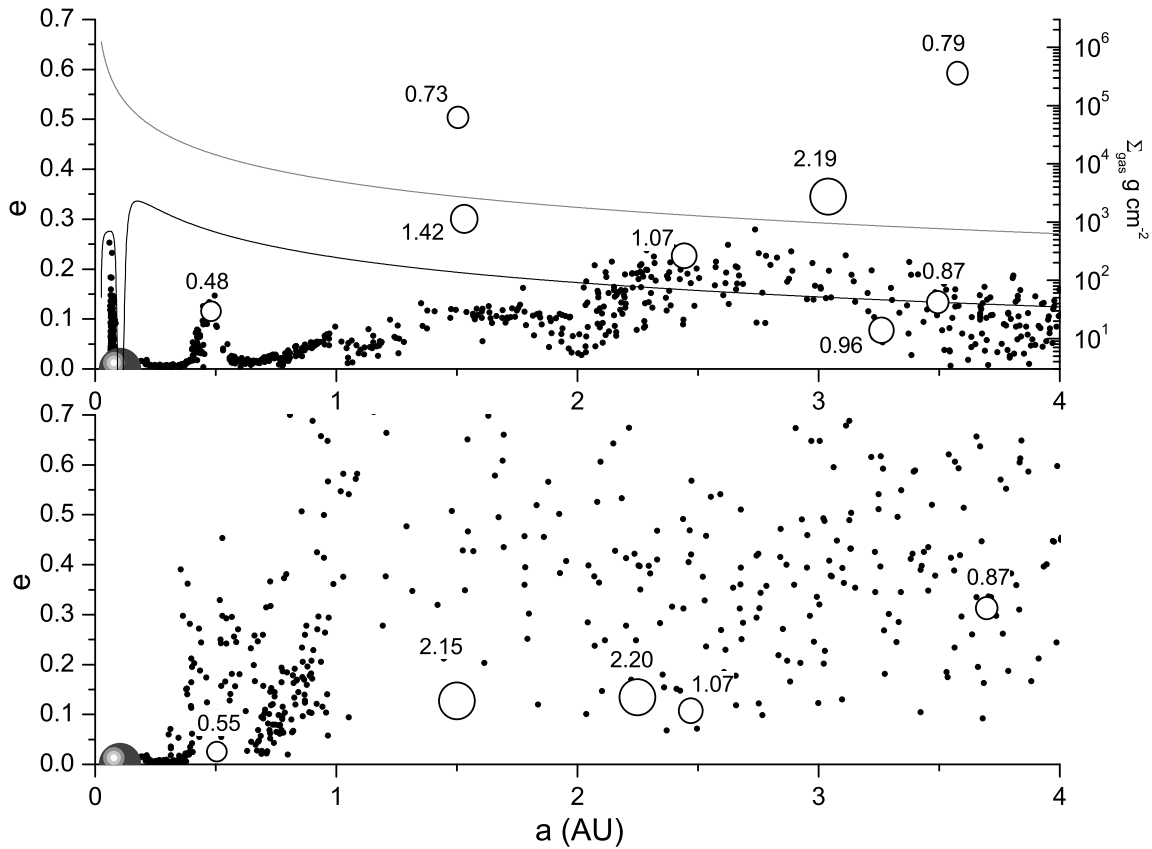}
 \caption{Eccentricities of bodies within 4~AU at the end of Scenario
  V (top panel) and after a further 2~Myr of gas-free accretion
  (bottom panel). Protoplanets are shown as white circles and are
  labeled with their mass in $\mathrm{m}_\oplus$. Super-planetesimals
  are indicated by black dots and the grey blob at 0.1~AU is the
  hot-Jupiter in its post-migration orbit. Gas density at the end of
  Scenario V is shown as the black curve in the top panel (read off
  the right-hand axis), the grey curve being the unevolved profile
  at $t = 0$}
 \label{figure:12}
\end{figure*}
Fig.~\ref{figure:12} shows the region between 0 -- 4~AU for the
scattered disk of Scenario V at the point when the giant planet
reaches 0.1~AU (top panel). Protoplanets of substantial size have
been scattered into the external disk. The result after a further
2~Myr of gas-free accretion is shown in the bottom panel. Dynamical
friction is sufficient to cause general damping of protoplanetary
orbits at the expense of excitation of the planetesimals. A
significant accretion event has been a giant impact and merger
between the 1.42 \& 0.73~$\mathrm{m}_\oplus$ protoplanets at
$\sim$1.5~AU resulting in a 2.15~$\mathrm{m}_\oplus$ body at $a =
1.47$~AU with $e = 0.13$. This planet, which lies within the
`maximum greenhouse' habitable zone, no longer crosses the orbit of
any of its neighbors and is probably a long-term survivor. There has
been some rearrangement of the intersecting protoplanetary orbits
beyond 2~AU with two protoplanets moving outward and circularizing
at $a > 4$~AU, off the right hand side of the figure, to compensate
for the 0.79~AU inward movement of a massive 2.2~$\mathrm{m}_\oplus$
planet, now at $a = 2.25$~AU with $e = 0.14$. This planet is
currently crossing the orbit of a smaller 1.07~$\mathrm{m}_\oplus$
body, the long-term fate of which will probably be accretion by one
of its two larger neighbors or incremental scattering by them to a
safe distance. The emergence of a stable terrestrial planetary
system from such a configuration seems highly probable, and this
result provides a clear prediction that terrestrial planets will be
found in the habitable zones of hot--Jupiter systems.

What will be the physical nature of such planets ? The issue of disk
and protoplanet composition after migration is discussed in detail
in Sect.~\ref{mixing}. Here we note, however, that the 2.15 and 2.20
m$_{\oplus}$ planets observed in Fig.~\ref{figure:12} are composed
of $\sim 30\%$ and $45\%$ of material originating from beyond the
snowline, respectively. Assuming that trans--snowline planetesimals
and protoplanets contain $75 \%$ water, a naive prediction is that
these planets will contain between 20 -- 30 \% water by mass. The
current water inventory of the Earth is estimated to be about 0.05
\% by mass. Evidently terrestrial planets forming in the habitable
zones of hot--Jupiter systems are likely to host deep global oceans
-- essentially being `water-worlds' \citep{kuchner,leger,raymond4},
even if significant loss of volatiles occurs during high--impact
accretion.

This prediction about water content, however, depends on the giant
planet forming out beyond the inner edge of the snowline, rather
than at its inner edge. The closer to the inner edge the giant
forms, the smaller the amount of volatile-rich material that will be
shepherded inward, and the lower the degree of volatile enrichment
experienced by terrestrial planets forming after migration of the
giant.

\subsection{Migration-driven compositional mixing.}\label{mixing}

\begin{figure}
 \resizebox{\hsize}{!}{\includegraphics{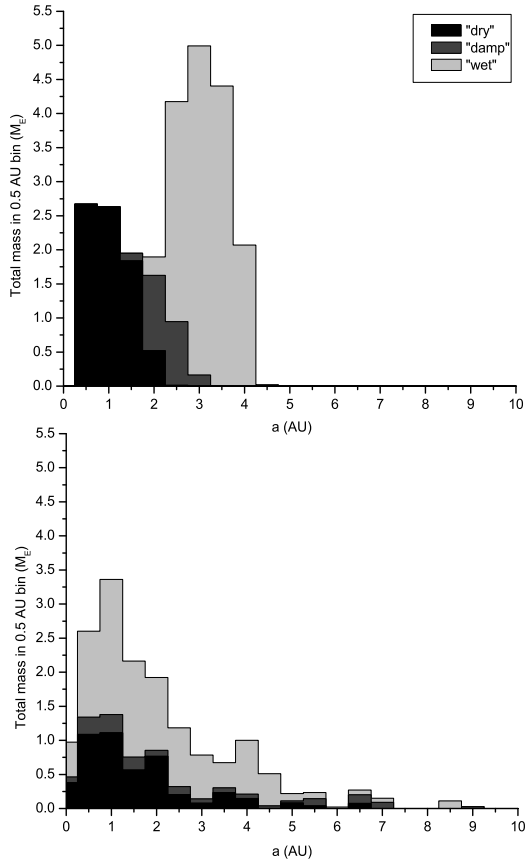}}
 \caption{Composition of the original solids disk in Scenario I
 (top panel) compared with the composition of the scattered disk
 generated through giant planet migration (bottom panel). The key
 is explained in the text.}
 \label{figure:13}
\end{figure}

A result found in all Scenarios I--V is homogenization of the solids
disk composition. This mixing occurs as the giant planet drives
material inward from the outer part of the disk whilst generating
the external disk via random scattering. An example is given in
Fig.~\ref{figure:13} which shows, for Scenario I, the composition of
the solids disk before migration (top panel) and that of the
surviving external disk after migration (bottom panel). In each case
the total mass of solid material is plotted in 0.5~AU width bins
with the histograms labeled ``dry", ``damp" and ``wet" representing
rocky material originating at $<$~2~AU, material characteristic of
chondritic meteorites between 2 -- 2.7~AU, and trans-snowline
material at $>$~2.7~AU, respectively. It is evident that a large
amount of material from beyond the snowline is shepherded into the
inner regions before being left behind. Compositional mixing is
similar in other scenarios, although a little less smooth in
Scenarios IV \& V which have had more time to accrete substantial
protoplanets from local material before the appearance of the giant
planet.

In contrast to other studies of water delivery to terrestrial
planets
\citep[e.g.][]{morbidelli,chambers2,raymond1,raymond2,raymond3,raymond4},
we do not assign an actual water mass fraction to our compositional
phases. This is because our simulations do not extend to the
completion of planetary accretion and our MMSN-type original solids
surface density profile, which includes a large step-increase at
2.7~AU (see Eq.~\ref{solids}), is different from those adopted in
the above-cited papers making a detailed comparison difficult.
Typically, we find that at the end point for early-migration
scenarios (I -- III) originally dry protoplanets that have found
their way into the scattered disk contain 0 -- 25\% of
trans-snowline material, but are surrounded by a large quantity of
volatile-rich planetesimals from which to accrete further. In later
scenarios (IV -- V), where the surviving protoplanets are more
mature, this range increases to 5 -- 70\%, and whilst there is less
remaining mass in small bodies to sweep up in these systems, the
accretion of volatile-rich material is not yet complete.

In general we can predict that the final terrestrial planets that
form in hot-Jupiter systems are likely to be much more volatile-rich
than the Earth, those in the habitable zone ending up with $> 10$\%
of their material originating from beyond the snowline. Accretion in
the external disk, however, may pass through an early high velocity
phase before completion with some protoplanets potentially losing
much of their volatile inventory during giant impacts
\citep{asphaug,canup}. Since water-rich planetesimals are abundant
throughout our external disks, however, protoplanets that have lost
their original volatile endowment in a catastrophic event should be
able to re-acquire some water before planetary accretion is
complete. Dry terrestrial planets appear improbable in hot-Jupiter
systems whereas Earth-like worlds and planets with deep global
oceans \citep{kuchner,leger,raymond4} may well be commonplace.

\section{Discussion.}\label{discussion}

The models presented here and in Paper I have only explored a small
region of the parameter space relevant to this problem, and have
inevitably adopted assumptions that simplify or omit potentially
important physical processes. We discuss some of these issues below
and their possible implications for our results. \\
({\it i}) {\em Giant planet mass \& radius.} \hspace{1mm} In all
simulations we have adopted a giant planet mass of
$0.5~\mathrm{M_J}$, with a density of $1~\mathrm{g~cm^{-3}}$ and
a radius of $0.85~\mathrm{R_J}$. Whilst the giant planet can accrete
solid matter as it migrates inward, it does not accumulate
any more gas and so its mass changes by $\lesssim$~5\% during
a run. Hot-Jupiters, however, come in a variety of masses
and although their average $m \sin i$ is less than a Jupiter mass,
more massive examples are known. We have not
yet tested the effect of increasing the mass of our migrating giant
planet, but expect that this would result in an enhanced and more
widespread scattering of material into the external disk, and
more accretion by the giant. \\
We have assumed a radius for our giant planet that is
representative of a fully contracted state, which may not be
realistic so soon after its formation.
Giant planets of approximately Jovian mass contract to radii
$\sim 2$ -- 3 R$_{\rm J}$ during their rapid gas accretion phase
\citep{pap_nelson05}, before cooling and contracting toward the
Jovian radius over longer time scales.
Adoption of a larger radius would probably result in greater
accretion of solids by the giant, with gas--drag--sensitive
planetesimals being preferentially accreted. \\
({\it ii}) {\em Migration halting mechanisms.} \hspace{1mm} The
reason why hot-Jupiters cease their inward migration close to the
central star and come to rest at $\lesssim$~0.1~AU is unknown.
Various mechanisms have been suggested, such as the giant planet
moving into a central magnetospheric cavity in the gas disk
\citep{lin2}, or fortuitous dispersal of the gas disk due to
photoevaporation \citep{trilling,armitage1,alibert}. We do not rely
on any particular mechanism here and merely note that if the former
operates then any of our scenarios may be appropriate, whereas if
the latter is true then our latest scenarios (IV \& V) are more
likely to be realistic as they operate at a time when most of the
gas disk has already been lost. In either case, we expect an
external disk to be generated during giant planet migration and
renewed terrestrial planet formation to follow. The dynamical state
of this scattered disk, however, will be affected by
photoevaporation of the gas disk as the giant migrates. The effect
of this on terrestrial planet formation during and after giant
planet migration will be
examined in a future paper. \\
({\it iii}) {\em Type I migration.} \hspace{1mm} We have not
included the effect of type I migration
\citep{ward2,tanaka2,tanaka3}, which would operate most effectively
on sub-gap opening bodies of $\gtrsim 1 \mathrm{m}_\oplus$, in any
of our published models to date and have assumed that it does not
play a major role. However, we discussed in detail its possible
influence in Paper I and confine ourselves here to the further
comment that even if type I migration operates with a reduced
efficiency to that expected from the canonical model, which seems
likely, it might still have a significant effect because of the
sensitivity of our model's outcome to the dissipation exerted on
solid bodies. Type I interactions would exert damping forces on
protoplanets additional to dynamical friction. This could have the
effect of reducing the eccentricity excitation of protoplanets
captured at resonances with the giant planet, hence reducing their
tendency to intersect the orbit of the giant and be scattered into
the external disk. The overall effect might be to increase the
shepherding effect of the giant planet at the expense of scattering,
enhancing the fraction of surviving material found interior to the
giant planet at the end of the simulation. If this material avoids
migration into the central star, it is possible that inclusion of
type I migration forces will assist in the formation and survival of
hot-Neptune and/or hot-Earth systems.
We are currently running simulations to examine this possibility.\\
({\it iv}) {\em Planetesimal size evolution.} \hspace{1mm} For
computational simplicity we have assumed a uniform planetesimal
population with radii of 10 km. In reality there would be a
distribution of planetesimal sizes determined from a balance between
their rates of accumulation and fragmentation. In regions of the
disk that are dynamically cold, the mean planetesimal size would
grow via binary mergers, whilst in regions that are more dynamically
excited destructive collisions could result in the planetesimal
population being ground down into smaller fragments. Fragmentation
of the planetesimal population in the context of the oligarchic
growth regime however is not necessarily an obstacle to planet
formation and may actually assist planetary growth by supplying
protoplanets with a more strongly damped feedstock, enhancing the
effect of gravitational focussing \citep{chambers3}. A particularly
challenging environment for planetesimal survival in our simulations
in in the compacted region of the disk between the 4:3 and 2:1
resonances with the giant planet. In this region, planetesimal
surface densities are enhanced (see the left hand panel in
Fig.~\ref{figure:8}) and the population as a whole is strongly
stirred generating eccentricities as high as 0.2 -- 0.3 (see
Figs.~\ref{figure:5} and \ref{figure:6}). Random velocities of
several km per second, far in excess of planetesimal escape
velocities, are indicated and mutual planetesimal collisions would
result in fragmentation. What this means for our model is unclear,
but it is possible that the fragment population, which would be much
more strongly affected by gas drag, would evolve inward rapidly,
thereby escaping the dynamically excited zone. It could then be
gathered efficiently by protoplanets and accreted. We speculate that
the overall effect of this process might be to reduce the
planetesimal fraction of the external disk material, but not to
reduce its overall mass as protoplanets should be scattered there
with the same, or even greater, efficiency.\\
({\it v}) {\em Eccentric giant planet.} \hspace{1mm} In this paper
we have assumed that the giant planet is on a high circular orbit as
it migrates. \citet{nagasawa} have shown that the interior disk can
be pre-stirred through the action of a sweeping secular resonance if
the giant is on a modestly eccentric orbit. This would modify the
accretion history of the inner disk.

\section{Conclusions.}\label{conclusions}

In this paper we have presented the results of simulations that
model terrestrial planet formation during and after the migration of
a gas--giant planet to form a 'hot--Jupiter'. This work is an
extension of our previous work \citep{fogg}, with improvements being
made by modelling the viscous evolution of the gas disk,
gap--formation and inner cavity formation in the gas disk due to the
gravitational influence of the planet, and self--consistent type II
migration. A popular belief has been that hot--Jupiter systems are
unlikely to host terrestrial planets, as migration of the giant
planet through the terrestrial planet zone was expected to sweep
that region of planet forming material. We find, however, that the
majority of this mass survives the migration episode as an interior
or exterior disk remnant from which terrestrial planet formation can
resume. This occurs via a combination of shepherding of the original
solids disk ahead of the giant, and random scattering of the
majority of this compacted material into orbits external to the
giant. The net effect is not a disappearance of planetary building
blocks from the inner system, but rather a stirring and mixing of
material originally formed at different
radial distances.

Now that more realistic gas dynamics have been included in our
model, generating partial cavity formation close to the central
star, gap formation in the vicinity of the giant planet's orbit,
consistent type II migration rates, and a decline in the overall
mass of gas present with time, we have found the following
qualitative differences between the results presented here and in
our previous work:

\begin{enumerate}
\item Scattering is favored over shepherding irrespective of the
maturity of the inner solids disk at the epoch of giant planet
migration. This occurs because of the rapid loss of gas from the
disk interior to the giant, and in the vicinity of the giant
planet's orbit, which reduces the damping of planetesimal
trajectories. The efficiency of shepherding and dynamical friction
is therefore reduced, increasing the probability that a given body
will come close enough to the giant to be scattered into an external
orbit. In all our scenarios $\gtrsim$~60\% of the original solids
disk material survives in a regenerated external disk.
\item The principal mass loss mechanism is accretion by the giant
planet, rather than the central star. This is because gas densities
close to the central star are reduced by 2 -- 3 orders of magnitude,
suppressing the gas-drag-induced orbital decay of planetesimals and
allowing the giant planet to catch up with and sweep through even
the innermost solids disk material. Most accretion by the giant
occurs at late times when scattering by the giant becomes less
effective due to the system being contained deep within the star's
gravitational potential well.
\item Hot-Neptune and/or hot-Earth survival is not favored
in our new model because
of the enhanced tendency of the giant planet to scatter material
outward and to accrete material at late times during the migration.
Accelerated protoplanetary accretion within the compacted shepherded
portion of the disk is observed, with objects growing to several
$\mathrm{m}_\oplus$, but their orbits typically destabilize at late
times due to capture in eccentricity--pumping mean motion resonances,
resulting in eventual collision with the giant planet.
However, due to the sensitivity of our results to levels of
dissipation close to the central star at late times,
and the potential influence of type I migration and eccentricity
damping that we have neglected,
it is premature to rule out the  possibility that hot--Neptunes
or hot--Earths can form and survive interior to Jovian planets
during their migration to become hot--Jupiters.
\item Whilst we predicted terrestrial planet formation in
hot-Jupiter systems from the results of our previous model, the
external disks generated by our new model appear to be even more
benign places for this to occur. These disks contain a greater mass
of material, are less dispersed, and are composed of a higher
fraction of small bodies capable of exerting dynamical friction and
hence damping the excited orbits of scattered protoplanets. The
formation of terrestrial planets of masses in the range $1 \le m_p
\le 3$ m$_{\oplus}$ occurred in or near the habitable zone during a
simulation that we continued after the giant stopped migrating. The
radial mixing of volatile--rich material from beyond the snowline
means that terrestrial planets forming in the habitable zones of
hot--Jupiter systems are likely to be ``water--worlds", hosting
deep, global oceans.

\end{enumerate}

The results presented in this paper make a clear prediction
that terrestrial planets will eventually be discovered
in the habitable zones of hot--Jupiter systems.
Such systems may be detectable by forthcoming missions
such as KEPLER, DARWIN and TPF.



\bibliographystyle{aa}

\listofobjects

\end{document}